\documentclass[12pt,preprint]{aastex}

\newcommand{\thetavec}{\mbox{\boldmath$\theta$}}
\newcommand{\alphavec}{\mbox{\boldmath$\alpha$}}
\newcommand{\uvec}{\mbox{\boldmath$u$}}
\newcommand{\svec}{\mbox{\boldmath$s$}}
\newcommand{\Svec}{\mbox{\boldmath$S$}}
\newcommand{\rvec}{\mbox{\boldmath$r$}}
\newcommand{\aprvec}{\mbox{\boldmath$r'$}}
\newcommand{\Rvec}{\mbox{\boldmath$R$}}
\newcommand{\Rvecone}{\mbox{\boldmath$r_1$}}
\newcommand{\nablavec}{\mbox{\boldmath$\nabla$}}

\begin{document}

\title{Microlensing under Shear}
\lefthead{Park & Ryu} \righthead{Microlensing under the Shear}

\author{Yoon-Hyun Ryu and Myeong-Gu Park}
\affil{Department of Astronomy and Atmospheric Sciences, Kyungpook
National University, Daegu 702-701, Korea}

\email{yhryu@knu.ac.kr, mgp@knu.ac.kr}

\begin{abstract}
Over two thousand Galactic microlensing events have been discovered
so far. All of them can be explained by events caused by  single or
multiple lenses (including binaries and planetary companions).
However, when a microlensing event occurs in highly dense star
fields such as in the Galactic bulge or in a globular cluster, it is
necessarily affected by the shear from the global distribution of
mass near the lens star. We investigate the distortions due to this
shear in the microlensing light curves and in the astrometric
microlensing centroid shift trajectories. As expected, the light
curve deviation increases as the shear increases and the impact
parameter decreases. Although the light curve in the presence of a
small shear is similar to the simple Paczy\'nski curve with a
slightly smaller impact parameter, the detailed difference between
the light curve with and without shear reflects the direction and
the magnitude of the shear. The centroid shift trajectory also
deviates from a simple ellipse in the presence of shear. The
distortion of the centroid shift trajectory increases as the impact
parameter decreases, and the shape of the trajectory becomes
complicated when the impact parameter becomes small enough. The
magnitude of the maximum distortion depends on the magnitude and the
direction of the shear. For a source trajectory in a given
direction, the time of the maximum distortion depends mostly on the
impact parameter and hardly on the shear. It is possible to
determine the magnitude of the shear and its direction if both the
time and the magnitude of the maximum astrometric distortion are
measured. The magnitude of the shear produced by the Galactic bulge
or a globular cluster falls in the range \(10^{-6}\)--\(10^{-4}\) in
normalized units. Although the actual determination of the shear
from the Galactic sub-structures will not be easy due to
complications such as binary companion, future large scale
microlensing experiments may enable us to determine the shear in
some high amplification events, leading eventually to mapping the
Galactic mass distribution.
\end{abstract}


\section{Introduction}
Since Paczy\'nski (1986; see also Gott 1981) first proposed
gravitational microlensing as a tool to detect massive astronomical
compact objects (MACHOs) in the Galactic halo, three groups (OGLE,
Udalski et al. 1992; EROS, Aubourg et al. 1993; MACHO, Alcock et al.
1993) independently discovered the first microlensing events, and
subsequent observations detected by now more than two thousand
microlensing events.

Gravitational microlensing has been applied to various fields of
astronomy, such as the studies of Galactic structure and stellar
populations and the search for extra-solar planets.   With
increasing potential of microlensing as a versatile astrophysical
tool, more advanced microlensing experiments such as highly precise
follow-up observations and pixel lensing observations are currently
being carried out, and next generation microlensing experiments such
as astrometric microlensing observations by using the \textit{Space
Interferometry Mission} $(\textit{SIM})$ and Keck and VLT
interferometers have been proposed.

Microlensing experiments are conducted toward very dense star fields
such as the Galactic bulge and the Magellanic clouds. When a
microlensing event occurs in these crowded fields, it is necessarily
affected by the  shear caused by the global distribution of mass
around the lens star. \citet{cha79,cha84} have discussed the effect
of a star on the macro-lensed image produced by the galaxy as a
whole, often referred to as the `Chang $\&$ Refsdal lens'. Their
lens model is characterized by `convergence' and `shear': the former
depends on the mass density within the beam and determines the
magnification of the image, while the latter depends on the mass
distribution outside of the beam and determines the distortion of
the image \citep[see e.g.][]{sch92}. They pointed out that the
configurations and the observational characteristics of macro-lensed
images can be significantly affected by a single star. However,
their work considered quasar lensing under shear, and focused only
on the image configurations. In this paper, we investigate how the
shear affects the stellar microlensing light curves and the
astrometric centroid shift trajectories.

The paper is organized as follows. We first introduce a single
point-mass microlensing in $\S$2 to compare with the point-mass
lensing with shear in $\S$3.1. In $\S$3.2 and $\S$3.3, we discuss
the distortions in the micro-lensed light curves and centroid shift
trajectories in the presence of the shear. In $\S$4, we discuss
possible applications to Galactic microlensing experiments. In
$\S$5, we summarize the results and conclude.

\section{Single Point Mass Microlensing}
Microlensing occurs when a lensing mass passes very close to the
line of sight to a background source star. When the lensing mass is
a single point mass, the source star splits into two images. The
separation between the two images is of the order of the Einstein
radius, which is related to the physical parameters of the lens
system by
\begin{equation}
\theta_E=\sqrt{\frac{4Gm}{c^2}\frac{D_{ls}}{D_l D_s}}\;,
\end{equation}
where $m$ is the mass of the lens, $D_l$ and $D_s$ the distances
from the observer to the lens and source, respectively, and $D_{ls}$
the distance from the lens to the source.  The angular positions of
the images with respect to the lens are
\begin{equation}
\thetavec_\pm = \frac{1}{2}\:
\bigg(\uvec\pm\sqrt{u^2+4}\;\frac{\uvec}{u}\bigg) \theta_E,
\end{equation}
where
\begin{equation}
\uvec = \left(\frac{t-t_0}{t_E}\right)\hat{x}+u_0\hat{y}
\end{equation}
is the projected lens-source separation vector in units of the
$\theta_E$, $u_0$ the impact parameter, $t_0$  the time of maximum
amplification. The Einstein ring radius crossing time or the
Einstein time scale, $t_E$, is given by \citep[see e.g.][]{gou00},
\begin{equation}
t_E=\frac{\theta_E}{\mu_{rel}}\;,
\end{equation}
where
\begin{displaymath}
\theta_E=\sqrt{\frac{4Gm}{c^2}\frac{\pi_{rel}}{AU}},\;\;\;\pi_{rel}=\frac{1}{D_l}-\frac{1}{D_s}\;,
\end{displaymath}
$\pi_{rel}$ and $\mu_{rel}$ are the relative source-lens parallax
and proper motion, respectively. The magnification of each
individual image is given by
\begin{equation}
A_{0,\pm} =
\frac{1}{2}\:\left[\frac{u^2+2}{u\sqrt{u^2+4}}\pm1\right],
\end{equation}
where $A_{0,+}$ and $A_{0,-}$ are the magnification factors of the
major and minor images, respectively. The position of the center of
light (centroid) corresponds to the magnification-weighted mean of
image positions, i.e.
\begin{equation}
\thetavec_0 = \frac{A_{0,+}\thetavec_++A_{0,-}\thetavec_-}{A_0}
\end{equation}
with $A_0=A_{0,+}+A_{0,-}$ is the total magnification. The centroid
shift $\delta\thetavec_0$ is defined as the difference between the
image centroid $\thetavec_0$ and the unlensed source position
$\thetavec_{s,0}$, and is related to the lensing parameters by
\begin{equation}
\delta\thetavec_0
=\thetavec_0-\thetavec_{s,0}=\frac{\theta_E}{u^2+2}\;\uvec\;.
\end{equation}
The position of the centroid shift caused by a single point-mass
lensing follows an ellipse \citep{wal95,jeo99}, which is represented
by
\begin{equation}
\frac{x^2}{a^2}+\frac{y^2}{b^2}=1\;,
\end{equation}
where the $\textit{x}$ and $\textit{y}$ represent the centroid shift
parallel and normal to the lens-source transverse motion,
respectively.  The semi-major axis $\textit{a}$ and semi-minor axis
$\textit{b}$ depend on the impact parameter $u_0$ as
\begin{equation}
a=\frac{\theta_E}{2(u_0^2+2)^{1/2}}\;,\;\;b=\frac{u_0\theta_E}{2(u_0^2+2)}\;.
\end{equation}

\section{Microlensing under Shear}
\subsection{Lens Equation}
When a source is lensed by a point mass $\textit{m}$ plus planar
mass distribution, the lens equation becomes \citep[see
e.g.][]{sch92,an05,an06}
\begin{equation}
\svec=\rvec-\frac{\rvec}{r^2}-\alphavec(\rvec)\;,
\end{equation}
where the two-dimensional vectors $\rvec$ and $\svec$ are the
positions of the images and the unlensed source, respectively. The
vector $\rvec$ in the lens (image) plane is normalized by
$r_E=\theta_E D_l$ and the vector $\svec$ in the source plane by
$s_E = \theta_E D_s$. The scaled deflection angle $\alphavec(\rvec)$
due to the additional mass distribution in the lens plane is given
by the gradient of the deflection potential $\psi$:
\begin{equation}
\alphavec(\rvec)=\nablavec\psi(\rvec)\;,
\end{equation}
where
\begin{equation}
\psi(\rvec)=\frac{1}{\pi}\int_{R^2}d^2r'\sigma(\aprvec)\ln|\rvec-\aprvec|\;.
\end{equation}
Here $\sigma(\rvec)$ represents the surface mass density
$\Sigma(\rvec)$ normalized by the critical surface mass density
$\Sigma_{cr}$,
\begin{equation}
\sigma(\rvec)=\frac{\Sigma(\rvec)}{\Sigma_{cr}},\;\;\;\Sigma_{cr}=\frac{c^2D_s}{4\pi
GD_{l'}D_{l's}}\;,
\end{equation}
where \(D_{l'}\) is the distance from the observer to the planar
mass distribution, which is in general different from \(D_l\), the
distance to the point mass $m$, and similarly for \(D_{l's}\).

When the mass distribution consists of a stellar mass and a much
larger-scale extended mass distribution, the lens equation is
approximated by a point mass plus quadrupole lens model (Chang \&
Refsdal 1984; see also Kovner 1987),
\begin{equation}
\svec\;=\;\rvec\:-\:\frac{\rvec}{r^2}\:-\:\left(
\begin{array}{cc}\kappa+\gamma & 0\\
0 & \kappa-\gamma
\end{array}\right)\rvec\;.
\end{equation}
The quadrupole term is specified by the convergence $\kappa$ and the
shear $\gamma$. These quantities are the two-dimensional second
derivatives of $\psi(\rvec)$:
\begin{equation}
\kappa=\frac{\psi_{11}+\psi_{22}}{2}\;,\;\;\gamma=\sqrt{\gamma_1+\gamma_2}\;,
\end{equation}
where
\begin{equation}
\gamma_1=\frac{\psi_{11}-\psi_{22}}{2}\;,\;\;\gamma_2=\psi_{12}=\psi_{21}\;,
\end{equation}
and $\psi_{ij}$ is the partial derivative of $\psi(\rvec)$ with
respect to $r_i$,
\begin{equation}
\psi_{ij}\equiv\frac{\partial}{\partial
r_i}\;\frac{\partial}{\partial r_j}\;\psi(\rvec)\;.
\end{equation}
If we define the rescaled coordinates $\Svec$ and $\Rvec$ as
\begin{equation}
\Svec\;\equiv\;\frac{\svec}{\sqrt{|1-\kappa+\gamma|}}\;,\;\;\Rvec\;\equiv\;\sqrt{|1-\kappa+\gamma|}\:\rvec\;,
\end{equation}
equation $(14)$ becomes
\begin{equation}
\Svec\;=\;\varepsilon \left(
\begin{array}{cc}
\Lambda & 0\\
 0 & 1
\end{array}\right)\Rvec\:-\:\frac{\Rvec}{R^2}\;,
\end{equation}
where
\begin{equation}
\varepsilon\equiv
\texttt{sign}(1-\kappa+\gamma)\;,\;\;\Lambda\equiv\frac{1-\tilde{\gamma}}{1+\tilde{\gamma}}\;,
\end{equation}
and the reduced shear $\tilde{\gamma}$ is
\begin{equation}
\tilde{\gamma}\equiv\frac{\gamma}{1-\kappa}\;.
\end{equation}
For convenience, $\tilde{\gamma}$ will be referred simply as the
shear henceforth.

In order to solve the lens equation, we introduce the polar
coordinates $(R,\varphi)$ in the lens plane: $R_x\equiv R
\cos\varphi$ and $R_y\equiv R\sin\varphi $. Then equation (19)
becomes
\begin{equation}
S_x= R (\varepsilon\Lambda-R^{-2})\cos\varphi ,\;\;S_y=
R(\varepsilon1-R^{-2})\sin\varphi\;,
\end{equation}
which yields a fourth-order equation for $R^2$ \citep{sch92},
\begin{eqnarray}
\Lambda^2R^8-[\varepsilon2\Lambda(\Lambda+1)+S_x^2+\Lambda^2S_y^2]R^6\nonumber \\
+[\Lambda^2+4\Lambda+1+\varepsilon2(S_x^2+\Lambda
S_y^2)]R^4\nonumber \\
-[\varepsilon2(\Lambda+1)+S_x^2+S_y^2]R^2+1=0\;,
\end{eqnarray}
We solve equation (23) by Laguerre's method.

\subsection{Light Curve}
Equation (23) has either zero, two, or four real roots, each of
which corresponds to the position of the individual image. The
Jacobian matrix of equation (19) is
\begin{equation}
\frac{\partial\Svec}{\partial\Rvec}=\left(
\begin{array}{cc}
\varepsilon\Lambda+\frac{R_x^2-R_y^2}{R^4} & \frac{2R_xR_y}{R^4}\\
\\
\frac{2R_xR_y}{R^4} & \varepsilon1-\frac{R_x^2-R_y^2}{R^4}
\end{array}\right)\;,
\end{equation}
whose determinant is
\begin{equation}
\textrm{det}\left(\frac{\partial\Svec}{\partial\Rvec}\right)=
\frac{1}{R^4}[\Lambda(R_x^2+R_y^2)^2+\varepsilon(1-\Lambda)(R_x^2-R_y^2)-1]\;.
\end{equation}
The magnification of each individual image is
\begin{equation}
A_{\gamma,i}=\left|\textrm{det}\left(\frac{\partial
\svec}{\partial\rvec}\right)\right|^{-1}=\;\frac{1}{1-\kappa+\gamma}\left|\textrm{det}\left(\frac{\partial
\Svec}{\partial\Rvec}\right)\right|^{-1}\;.
\end{equation}
The total magnification is the sum of the magnifications of
individual images, $A_\gamma=\sum_{i}{A_{\gamma,}}_i$. In order to
quantify how much the light curve in the presence of shear deviates
from that in the absence of shear, we define the excess
magnification as
\begin{equation}
\delta A\equiv A_\gamma-A_0\;,
\end{equation}
where $A_\gamma$ and $A_0$ represent the magnifications with and
without shear, respectively. In Figure 1, we present the contour
maps of magnification $A_\gamma$ (left panels) and excess
magnification $\delta A$ (right panels) as a function of source
position ($s_x, s_y$) for $\tilde{\gamma}=10^{-2},$ $10^{-4},$ and
$10^{-6}$, respectively. The caustics appear as the central diamonds
in the left panel of Figure 1, whose full width on the
$\textit{x}$-axis is $4\gamma(1-\kappa+\gamma)^{-1/2}$ and that on
the $\textit{y}$-axis is $4\gamma(1-\kappa-\gamma)^{-1/2}$
\citep[][]{han05}. When the source is outside the caustic, i.e.
$u_0\gtrsim4{\gamma}$, the number of images is two as in the simple
lensing without shear.

The series solution of equation (23) can be derived under the
assumption $\kappa\ll\gamma\ll|\svec|\ll1$. The excess magnification
$\delta A$ is calculated in powers of \(\tilde\gamma\), and the
leading term yields
\[
  \delta A
  \simeq\tilde\gamma\left(-\frac{1}{2s}+\frac{3s_y^2}{s^3}\right).
  \]
Although other geometric configurations will yield different values,
the order of magnitude of the deviation will be the same, \( \delta
A \sim \tilde\gamma / u_0 \). This shows that even very small shear
can produce a significant deviation in the light curve if the impact
parameter is small enough, that is in high-magnification events.

We now investigate the light curves for typical source trajectories.
Since the trajectory of the source does not coincide with the
direction of the shear in general, we choose source trajectories
with various angles (see Fig. 2). Figure 2 shows the angle
$\vartheta$ defined as the angle between the $\textit{x}$-axis and
the source trajectory. The dotted ring around the lens (its position
marked by `$\times$') is the Einstein ring. In the left panel of
Figure 3, we present the lensing light curves for the corresponding
source trajectories marked in Figure 2 with values of the shear
$\tilde{\gamma}=$ 0.01, 0.005, 0.001, and $\tilde{\gamma}=$ 0.0 (no
shear). All light curves have the same impact parameter, $u_0= 0.3$.
The minimum magnification $A_\gamma$ is greater than the minimum of
$A_0$ by a factor $1/(1+\tilde{\gamma})(1-\tilde{\gamma})$ for
$\tilde{\gamma}<1$. This is because the brightness of the source is
increased by the shear alone even in the absence of the point-mass
lens. In the right panel of Figure 3, we present the magnitude and
pattern of $\delta A$ for various shear values and source
trajectories. We find that the excess magnification increases as
$\tilde{\gamma}$ increases and the excess becomes maximum when the
source trajectory is parallel to the shear direction ($\vartheta=
0^\circ$). Even if the light curves have the same shear and the
impact parameter, the positions and heights of the peak deviations
vary with $\vartheta$. We find that the value of maximum $\delta A$
decreases as source trajectory becomes perpendicular to the shear
direction ($\vartheta=90^\circ$). We also find that unless the
source trajectory is parallel or perpendicular to the shear
direction, the deviation $\delta A$ becomes asymmetric in the
presence of the shear.

In real observations of microlensing, however, each light curve will
be first fitted by a theoretical microlensing light curve such as
the Paczy\'nski curve. So we check how the light variation induced
by the presence of shear deviates from the single mass lensing
curve. In Paczy\'nski curve, the peak amplification $A_p$ is a
simple function of the impact parameter $u_0$,
\begin{equation}
A_p=\left[1-\left(\frac{u_0^2}{2}+1\right)^{-2}\right]^{-1/2}\;.
\end{equation}
We compare the light curve in the presence of shear with the
Paczy\'nski curve that has the same peak magnification $A_p$ at
$t=t_0$ against the background magnification due to the shear alone,
$A_{\gamma}|_{s\rightarrow\infty}$. The impact parameter $u_{0p}$
(eq. [28]) from the fitted Paczy\'nski curve is always slightly
smaller than the true impact parameter $u_0$ because shear increases
the maximum magnification. The detailed shapes of the deviations
from the Paczy\'nski curve for various $\tilde{\gamma}$ and
$\vartheta$ are shown in Figure 4. As expected from the Figure 3,
the light curve becomes asymmetric in the presence of the shear when
the source trajectory is not parallel or perpendicular to the
direction of the shear. By fitting the deviation from Paczy\'nski
curve with an appropriate lens model with shear, it is possible, in
principle, to determine the magnitude and the direction of the shear
if the photometric accuracy is good enough to measure the deviation.
In real observations, the light curve as a whole will be fitted by
the Paczyi\'nski curve, and the shape and the maximum value of the
deviation can be somewhat smaller than that from the simple fitting
of the maximum amplification.

\subsection{Centroid Shift}
We define the centroid shift by
\begin{equation}
\delta\thetavec_\gamma \equiv \thetavec_\gamma - \thetavec_s\;,
\end{equation}
where
\begin{equation}
\thetavec_\gamma= \frac{\sum
A_{\gamma,i}\;\rvec_i}{A_\gamma}\;\theta_E\;,\;\;\thetavec_s=\svec_\gamma\theta_E\;,
\end{equation}
$\svec_\gamma$ is the position of the image that would result from
the shear alone in the absence of the point mass, i.e.
$\svec_\gamma=\svec+\alphavec(\svec_\gamma)$. When microlensing
occurs in the presence of shear, astrometric observations will
measure $\delta\thetavec_\gamma$. From equations (18) and (22), we
see that
\begin{equation}
\svec_{\gamma,x}=\frac{s_x}{1-\kappa-\gamma},\;\;\;\;
\svec_{\gamma,y}=\frac{s_y}{1-\kappa+\gamma}\;,
\end{equation}
which shows the whole field is sheared even before stellar
microlensing occurs. This affects the observed proper motion of the
source as well as the estimate of the lensing parameters. If the
source proper motion increases by the factor $\Lambda^{-1}$, the
Einstein ring radius crossing time would change to
\begin{equation}
t_E'=\Lambda\;t_E\;.
\end{equation}

The trajectories of $\delta\thetavec_\gamma$ in units of $\theta_E$
are shown in Figure 5 for given $\vartheta's$ and
$\tilde{\gamma}'s$.  The left panel shows the change of centroid
shift trajectories depending on $\vartheta$ and $\tilde{\gamma}$.
The centroid shift trajectories ($\textsl{dotted, dot-dashed, and
dashed curves}$) deviate from a simple astrometric ellipse
($\textsl{solid curve}$), and both the shape and the magnitude of
the distortion vary with $\vartheta$ and $\tilde{\gamma}$. The right
panel of Figure 5 shows the variation of the centroid shift
trajectory depending on the impact parameter $u_0$. The trajectories
of centroid shift for different $u_0$ in the presence of the shear
are different from each other.  The shape becomes more complex and
the distortion increases as $u_0$ decreases. The distortion becomes
very large for small values of $u_0$ even when the shear is small.

To quantify the deviation of the centroid shift trajectory, we
calculate the excess centroid shift defined as
\begin{equation}
\Delta\thetavec\equiv\delta\thetavec_\gamma-\delta\thetavec_0\;,
\end{equation}
where $\delta\thetavec_\gamma$ and $\delta\thetavec_0$ represent the
centroid shifts with and without shear, respectively. The deviation
$\Delta\thetavec$ is really the deviation from the centroid shift
ellipse which is expected in a point mass lensing without shear. The
trajectories of $\Delta\thetavec$ are shown in the left panel of
Figure 6. The arrow in each panel shows the direction of the
centroid motion with the progress of time. All excess centroid
shifts have one twist. This is similar to the planet-induced
microlensing centroid shift \citep{han02}. The magnitude of the
excess centroid shift defined as
$\Delta\theta\equiv|\delta\thetavec_\gamma|-|\delta\thetavec_0|$ is
shown in the right panel of Figure 6. We find that major deviation
occurs when $-2<(t-t_0)/t_E<2$, as expected. The sign of
$\Delta\theta$ can be either positive or negative. The detailed
shape of $\Delta\theta$ as a function of $(t-t_0)/t_E$ depends on
$\vartheta$.   For intermediate values of $\vartheta$, e.g.,
$\vartheta=30^\circ$ or $60^\circ$, $\Delta\theta$ changes its sign
near $t=t_0$.

We further investigate the magnitude of the maximum distortion,
$\Delta\theta_{max}$, and the time of maximum distortion, $t_{max}$.
Figure 7a shows how $\Delta\theta_{max}$ in units of $\theta_E$
varies as a function of $\tilde{\gamma}$ for different values of
$\vartheta$ and $u_0$. For small enough value of $\tilde{\gamma}$,
$\log\Delta\theta_{max}$ increases linearly as $\log\tilde{\gamma}$
increases.   For a given $\tilde{\gamma}$, $\Delta\theta_{max}$
becomes maximum when $\vartheta=0^\circ$ and $90^\circ$ and minimum
when $\vartheta=45^\circ$. Figure 7b shows the dependence of
$\Delta\theta_{max}$ on $u_0$ for given $\tilde{\gamma}$ and
$\vartheta=45^\circ$. Figure 8a and 8b show $t_{max}$ in units of
$t_E$ as a function of $\tilde{\gamma}$ for different $\vartheta$
and $u_0$, which Figure 8c shows the dependence of $t_{max}$ on
$u_0$. For each $\vartheta$ and $u_0$, $t_{max}$ is nearly constant,
independent of $\tilde{\gamma}$. Since $u_0$ can be determined from
the centroid shift trajectory, $\tilde{\gamma}$ can be determined
from the $\Delta\theta_{max}$ (Fig. 7a) and $\vartheta$ from
$t_{max}$ (Fig. 8a). Therefore, it is possible to determine the
shear and its direction if we determine by astrometry both the time
$t_{max}$ and the magnitude of the maximum astrometric distortion
$\Delta\theta_{max}$. Needless to say, one can always fit the full
lensing (with shear) model to each individual case, and determine
the shear and its direction.

As in $\S3.2$, we can also calculate the series expressions for
$\Delta\theta$ in powers of $\tilde\gamma$ under the same
assumption:
\begin{equation}
\Delta\theta\simeq\tilde{\gamma}\left[2\left(\frac{s^2-1}{s}\right)
+4s_y^2\left(\frac{1-s^2}{s^3}\right)\right]
\end{equation}
where
\begin{equation}
s=\sqrt{T^2+u_0^2},\;s_y=-T\sin\vartheta+u_0\cos\vartheta,
\end{equation}
and
\begin{equation}
T\equiv\left(\frac{t-t_0}{t_E}\right).
\end{equation}
The time of maximum distortion, $t_{max}$, is given by a root of the
equation
\begin{equation}
T^5+(4u_0^2+1)T^3+(2u_0^3+4u_0)\tan2\vartheta\;
T^2+(3u_0^4-5u_0^2)T+(2u_0^5-2u_0^3)\tan2\vartheta=0.
\end{equation}
This equation has three real roots, and the one with the smallest
absolute value corresponds to $t_{max}$. The angle $\vartheta$ is
now
\begin{equation}
\vartheta=\frac{1}{2}\arctan\left[\frac{-t_{max}^5-(4u_0^2+1)t_{max}^3
+(5u_0^2-3u_0^4)t_{max}}{(2u_0^3+4u_0)t_{max}^2+(2u_0^5-2u_0^3)}\right].
\end{equation}
Hence, we can also approximately determine $\tilde\gamma$ and its
direction $\vartheta$ from equations (34) and (38) from $t_{max}$
and $\Delta\theta_{max}$.

\section{Application to Galactic Microlensing}
Now we discuss Galactic microlensing affected by the shear. Consider
a microlensing system that consists of a single lensing star under a
shear. The shear field can be produced by any Galactic
sub-structures such as globular clusters and the Galactic bulge.
Here, we only consider the Galactic bulge and globular clusters as
typical examples of the shear, and model their mass distribution as
a point mass or the Plummer's model. The Galactic bulge is
significantly extended along the line of sight and, therefore, the
mass distribution has to be weighted by the factor \( (D_{l'}
D_{l's})/D_s \) and projected along the line of sight; the bulge
mass distribution located near the source plane contributes little
while those near the half of the distance to the source contributes
most. But in this work we model the bulge as a planar mass
distribution at the same distance as that of the lensing star for
simplicity.

1. Point mass: When the lensed images are located far from the
source of the shear, the shear field may be approximated by that
produced by a point mass.   When the mass distribution consists of a
lens with mass $m$ located at the origin of the coordinate and an
additional mass with mass $M$ located at $\Rvecone$ (in units of
$r_E$), the (additional) deflection potential $\psi(\rvec)$ in
equation (12) becomes
\begin{equation}
\psi(\rvec)=\frac{M}{m}\ln|\rvec-\Rvecone|\;.
\end{equation}
Shear $\gamma$ and convergence $\kappa$ are calculated from equation
(15):
\begin{equation}
\gamma=\frac{M}{m}\frac{1}{R^2}\;,\;\;\kappa=0\;.
\end{equation}
Convergence disappears ($\kappa=0$) because the beam is empty, and
the reduced shear is equal to the shear, $\tilde{\gamma}=\gamma$.

2. Plummer's model: As an example of extended mass distribution, we
choose the Plummer's model that approximates the surface mass
distribution of a globular cluster \citep{plu15,bin87}. In the
Plummer's model, the surface mass density is expressed as
\begin{equation}
\Sigma(\rvec)=\frac{M}{\pi r_0^2}\frac{1}{[1+(r/r_0)^2]^2}\;,
\end{equation}
where $M$ is the total mass and $r_0$ is the core length in units of
$r_E$. The radius $r_h$ containing half the total mass is equal to
1.3048$r_0$. Then, the deflection potential $\psi(\rvec)$ due to the
Plummer's mass distribution centered at $\Rvecone$ becomes
\begin{displaymath}
\psi(\rvec)=\frac{1}{\pi}\int_{R^2}d^2r'\frac{\Sigma(\aprvec-\Rvecone)}{\Sigma_{cr}}\ln|\rvec-\aprvec|
\end{displaymath}
\begin{equation}
\;\;\;\;\;=\frac{M}{m} \ln[{(\rvec-\Rvecone)^2+r_0^2}]^{1/2}\;.
\end{equation}
Again, $\gamma$ and $\kappa$ are calculated from equation (15):
\begin{equation}
\gamma=\frac{M}{m}\frac{{r_1}^2}{({r_1}^2+r_0^2)^2}
\end{equation}
and
\begin{equation}
\kappa=\frac{M}{m}\frac{r_0^2}{({r_1}^2+r_0^2)^2}\;.
\end{equation}
Unlike the point mass case, the convergence $\kappa$ is not zero
because there is mass within the beam.

Figure 9a shows the values of $\tilde{\gamma}$ and $\kappa$ as
functions of the distance from the center of the mass distribution
when the shear is produced by the Galactic bulge with a total mass
of $M=1.3\times10^{10}$ $\textrm{M}_\odot$ and located at 8.5 kpc
from the Earth. We assume that the lens star of $1\textrm{M}_\odot$
is also located at 8.5 kpc and the source star at 9.5 kpc. Solid
curve represents $\tilde{\gamma}$ when the bulge is modeled as a
point mass while dotted curve as the Plummer's model with $r_0=500$
pc. Dashed curve shows $\kappa$ for the Plummer's model. Figure 9b
is the same as Figure 9a for a $10^6$ $\textrm{M}_\odot$ globular
cluster at $4$ kpc with $r_0=2$ pc, 1 $\textrm{M}_\odot$ lens star
at the same 4 kpc, and a source star at $8.5$ kpc. Hence, we expect
the shear $\tilde{\gamma}$ produced by typical globular clusters or
the Galactic bulge to be in the range of $10^{-6}\sim10^{-4}$.

Shown in Figure 10a is the maximum distortion of the centroid shift
in arcseconds when a microlensing event occurs in the Galactic
bulge. If future astrometric observation can achieve the positional
accuracy down to $\sim1$ micro-arcsec, then $\tilde{\gamma}\sim
10^{-5.5}$ shear field can be detected in very high magnification
events with $u_0\leq 0.002$. When a microlensing event occurs near a
globular cluster (Fig. 10b), $\tilde{\gamma}\sim 10^{-4.5}$ shear
field can be detected for events with $u_0\leq0.05$. On the other
hand, in order to detect the deviation of a typical light curve for
$\tilde{\gamma}=10^{-5.5}$ and $u_0=0.01$, photometric accuracy
should be better than $\Delta m \leq10^{-3.5}$.

Although we may be able to measure the shear in microlensing events,
the measurement alone does not tell us about the source of the
shear. The shear may be from the Galactic sub-structures, but it can
also be from many other objects. The most obvious source is the
binary companion. A binary companion at a typical distance of $\sim
35$ AU (Duquennoy \& Mayor 1991) will produce a shear close to
\(10^{-3}\), two to three orders of magnitude larger than the shear
by the Galactic bulge. The distribution of the binary period and the
mass ratio (Duquennoy \& Mayor 1991) implies 85\% of all binary
companions will produce \( \gamma \ge 10^{-6} \). So the shear due
to binary companion will dominate or be comparable to the shear
expected from Galactic sub-structures. The shear can be also
produced by planetary companions. Planets detected in the current
microlensing experiments produce shear in the range
$\tilde\gamma\sim10^{-3}-10^{-5}$. A typical Earth-mass planet at a
distance of $\sim2R_E$ ($R_E\simeq4$ AU for 1 $\textrm{M}_\odot$)
from the lens produces $\tilde\gamma\sim10^{-6}$, while Pluto-mass
planet at a distance $\sim 10R_E$ produces
$\tilde\gamma\sim10^{-10}$. Projected companion stars, unassociated
but located near the lensing star in the projected sky, will also
affect lensing similarly. Hence, in practice it will be difficult to
identify the shear by Galactic sub-structures against the shear by
companion stars. Still, there can be some microlensing events that
are not affected by companion star or in which the shear by Galactic
sub-structure may be measured from the statistical analysis of many
events as in cosmological weak lensing systems.

There may also be cases for which we may set an upper limit on the
value of the shear. A very small upper limit on shear suggests the
non-existence of a binary companion or a Galactic structure. For
example, if we measure \( \gamma < 10^{-7.2} \), we can expect that
the lens does not have a companion with more than 95\% confidence if
we assume the distribution of the binary companion by Duquennoy \&
Mayor (1991).

Complications in usual microlensing events, for example, blending
and binary source, will also affect microlensing under shear. For
example, blending decreases the deviation in the image centroid
shift and the light curve (Fig. 11b as compared to Fig. 11a) and the
binary source completely messes up the centroid shift trajectory
(Fig. 11c). However, since the shape of the trajectory mainly
depends on the size of the shear, fitting the full trajectory may
sort our these complications.

Although the typical shear expected from the Galactic bulge or
globular clusters is too small or the clear case that enables the
determination of shear is too infrequent to be comfortably detected
by current or near-future microlensing experiments, next generation
microlensing experiments may enable us to measure the magnitude and
the direction of even smaller shear fields among numerous
microlensing events. Then, from the theoretical framework of weak
lensing, the shear map can be inverted to reproduce the mass
distribution \citep[see e.g.,][]{mel99,bar01,ref03}, making it
possible to map or at least constrain the Galactic mass
distribution. Since all our discussions are based on normalized
units, the reproduced mass distribution will be in units of
\(\Sigma_{cr}\) per \(\theta_E^2\). If \(D_s\) and \(D_l\) are
additionally determined, the mass distribution can be determined in
physical units.

\section{Summary}
We investigated microlensing under a shear, which might be produced
by Galactic sub-structures such as globular clusters or the Galactic
bulge. We analyzed its effect on the microlensing light curves and
astrometric centroid shift trajectories. We found the followings:

1. The light curve deviation from the Paczy\'nski curve increases as
the shear increases and the impact parameter decreases. The
positions and heights of the maximum deviation vary depending on the
direction of the source trajectory: the light curve becomes
asymmetric if the source trajectory is not parallel or perpendicular
to the shear direction.

2. The centroid shift trajectory in the presence of shear deviates
from a simple ellipse, especially when the source is within the
Einstein ring. The magnitude of the maximum distortion depends on
shear and its direction, and becomes largest when the trajectory is
parallel or perpendicular to the shear direction. The time of
maximum distortion is nearly independent of the impact parameter.
The distortion of the centroid shift trajectory increases as the
impact parameter decreases and the shape becomes very complex when
the impact parameter is very small.

3. If we measure the distortion of the astrometric centroid shift
trajectory and the time of the maximum distortion, we can determine
the shear and its direction.

4. The magnitude of the shear produced by the Galactic bulge or
globular clusters near the Galactic center is of the order of
$10^{-6}$ to $10^{-4}$ in normalized units. This shear, in
principle, could be detected by future microlensing experiments,
especially in high magnification events.

Successful measurement of the shear in various directions in the
Galaxy with next generation microlensing experiments may eventually
lead to the mapping of the Galactic mass distribution.

\acknowledgements We thank the referee, Scott Gaudi, for many
insightful comments, which have improved the paper greatly. We also
thank Dong Wook Lee and Cheongho Han for careful reading of the
manuscript. This work is the result of research activities
Astrophysical Research Center for the Structure and Evolution of the
Cosmos (ARCSEC) supported by the Korea Science \& Engineering
Foundation (KOSEF).

\clearpage
\begin{figure}
    \epsscale{1.0} \plotone{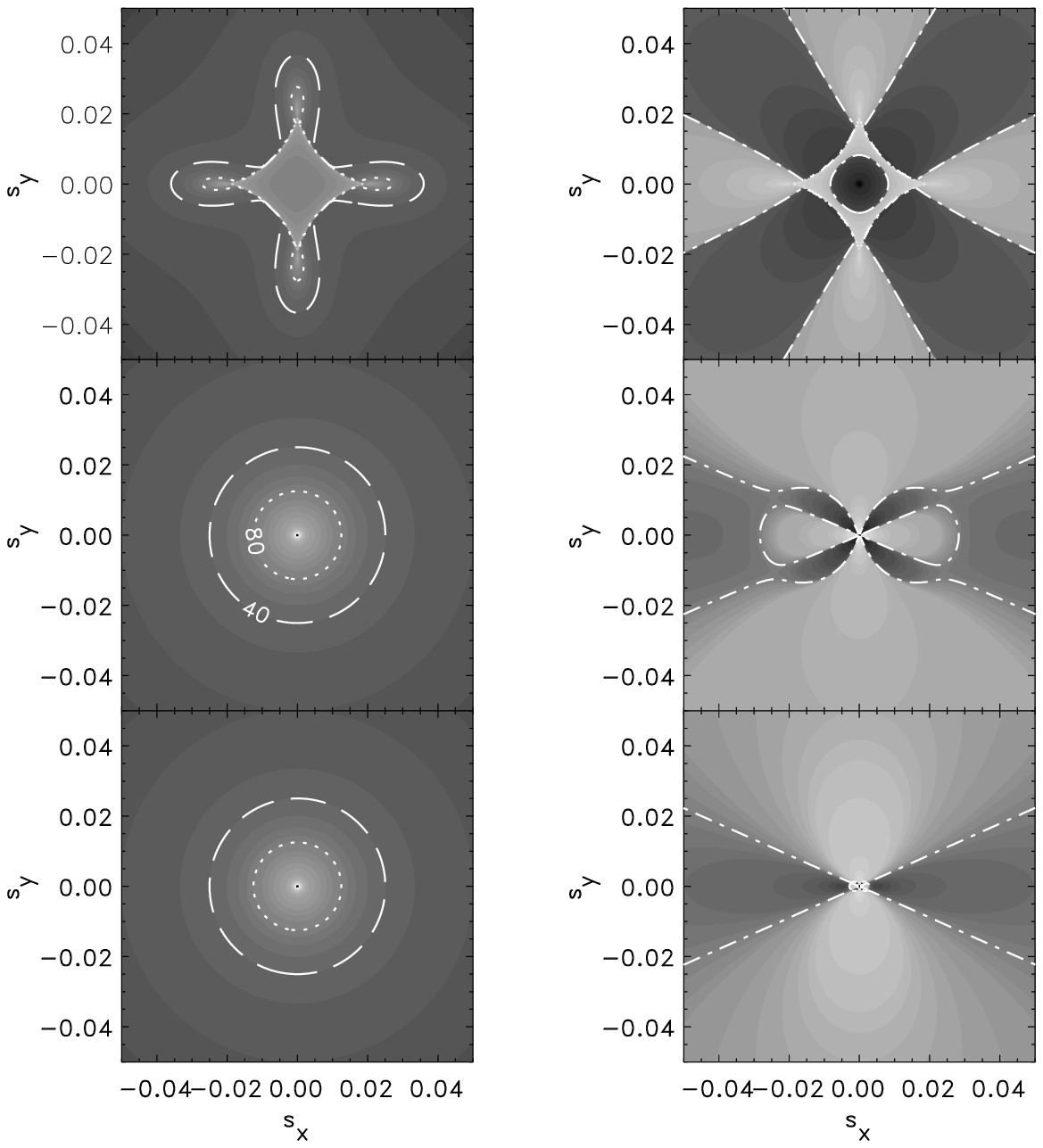}
    \caption{Magnification map. Grayscale maps of the magnification $A_\gamma$ (left panel)
    and the excess magnification $\delta A$ (right panel) as
     a function of source position ($s_x, s_y$) for reduced shear $\tilde{\gamma}=10^{-2}$ (top),
      $10^{-4}$ (middle), and $10^{-6}$ (bottom). Contours in the left
     panels represent $A_\gamma=40$ (long dashed curve) and
     $A_\gamma=80$ (dotted curve). Gray scale in the right panels represents positive (bright) and negative (dark)
     deviation regions and dot-dashed curve shows $\delta A=0$ regions.}
\end{figure}

\begin{figure}
    \epsscale{1.0} \plotone{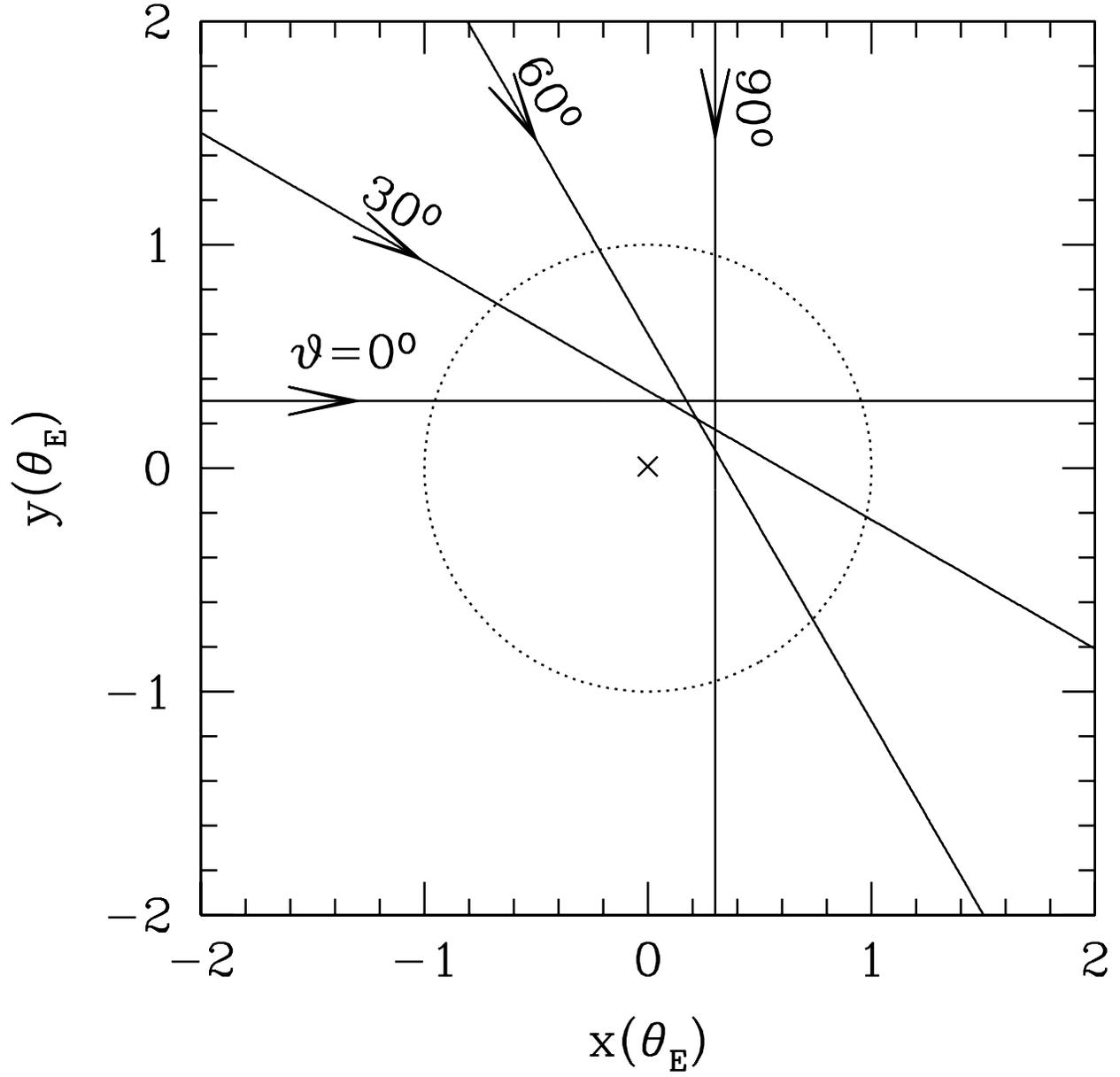}
    \caption{Geometry of lensing.   The straight lines with an
    arrow represent source trajectories with different $\vartheta$.
    We define $\vartheta$ as the angle between the $\textit{x}$-axis and
    the source trajectory. All source trajectories have the same impact
    parameter $u_0 = 0.3$.   The dotted ring around the lens, marked
    by `$\times$', shows the Einstein ring.}
\end{figure}

\begin{figure}
    \epsscale{1.0} \plotone{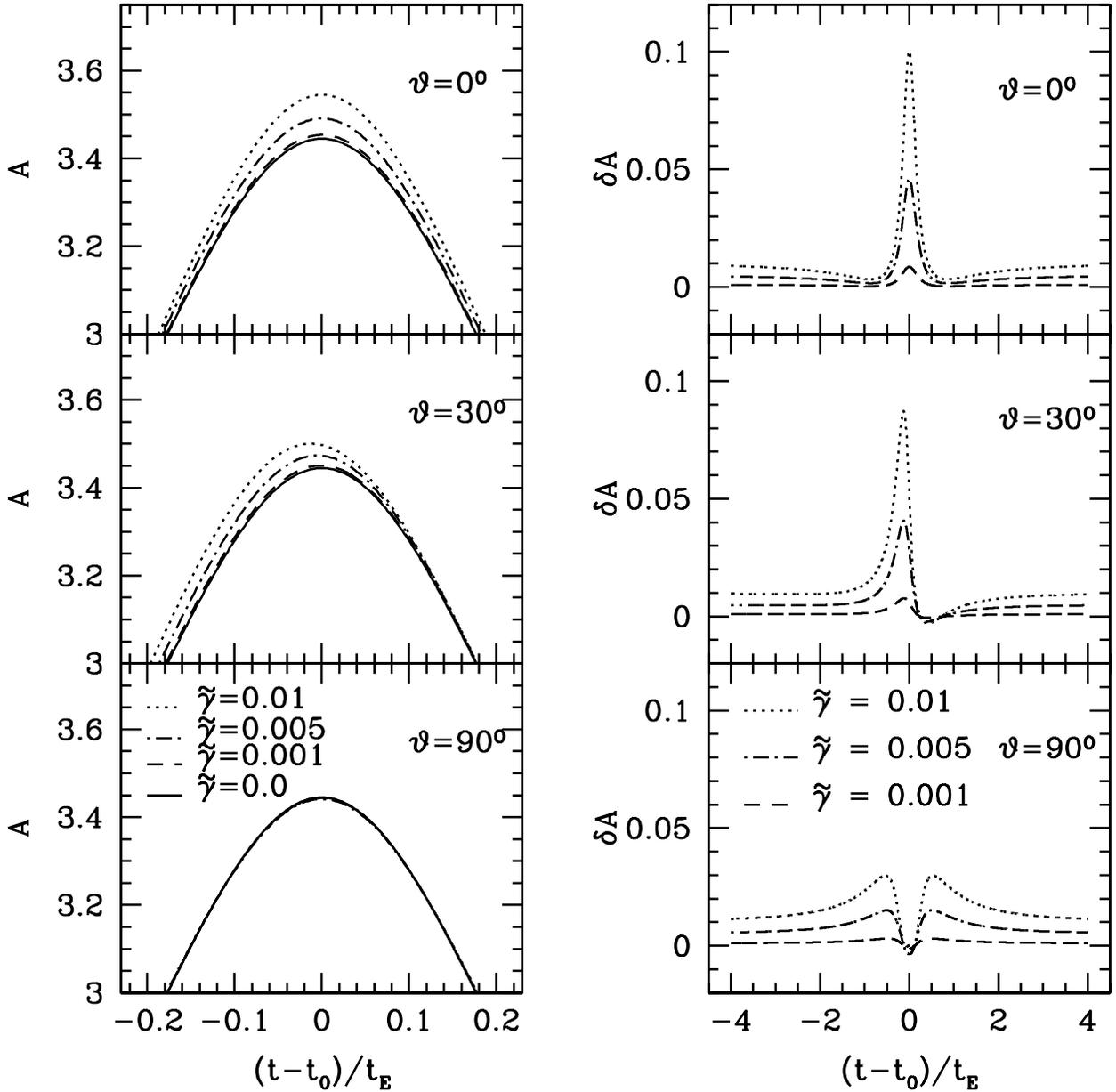}
    \caption{Light curves and the excess magnifications.
     The left panels show the light curves for the source trajectories ($u_0=0.3$) in Figure
    1 with the shear value of $\tilde{\gamma}=0.01, 0.005, 0.001$.   All curves
    show the magnification near the peak, and the solid curves are
    the magnification without shear $(\tilde{\gamma}=0.0)$.
    The right panels show the difference in magnification between the light curve in the
    presence of shear and that in the absence of shear as a function of
    time for three different source trajectories.}
\end{figure}

\begin{figure}
    \epsscale{1.0} \plotone{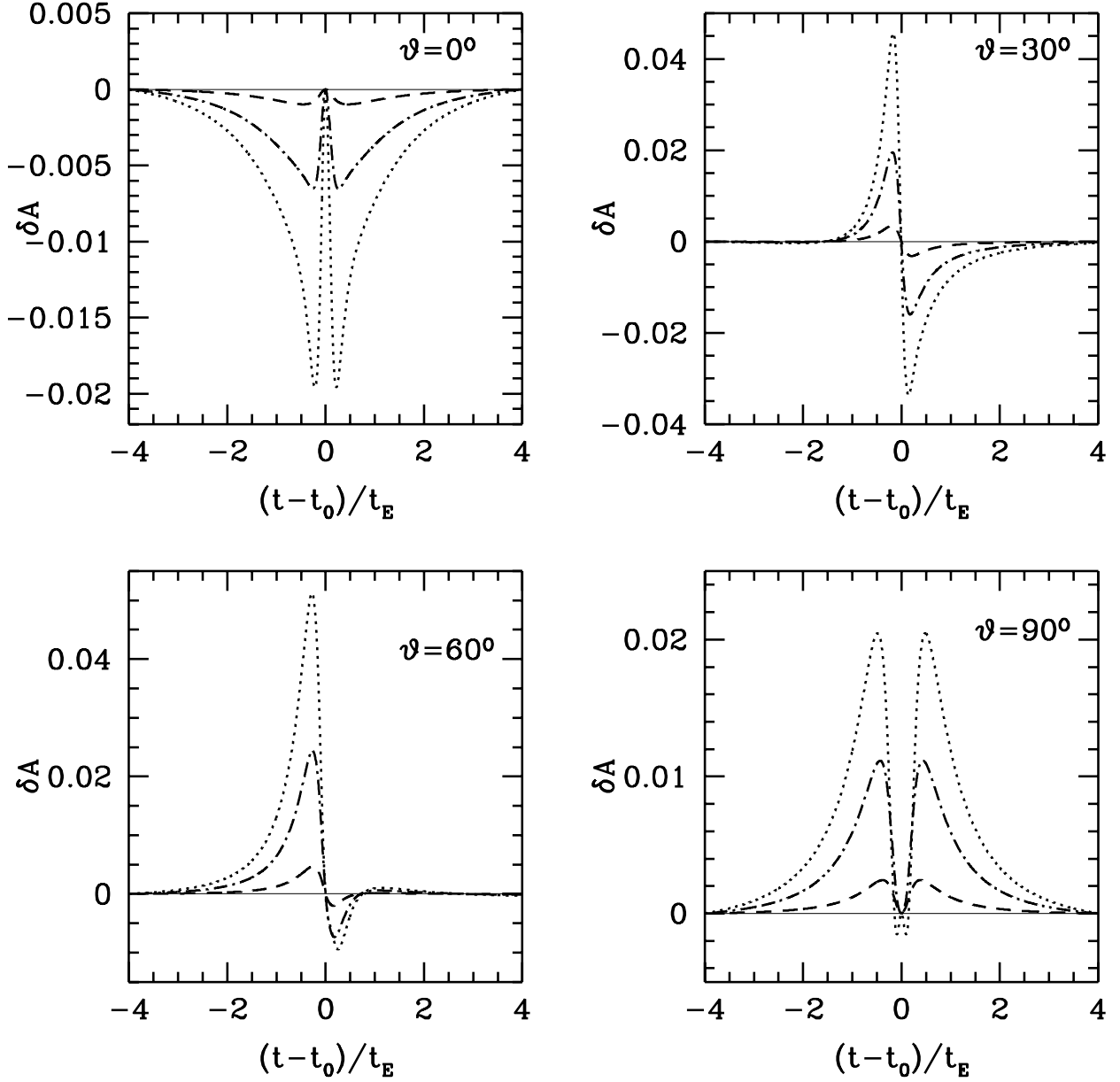}
    \caption{Deviation in the light curves.
    Difference between the light curve with shear and the corresponding Paczy\'{n}ski curve
    as a function of time for different source trajectories.
    Lines are for the same $\tilde{\gamma}'s$ in Figure 3.}
\end{figure}

\begin{figure}
    \epsscale{1.0} \plotone{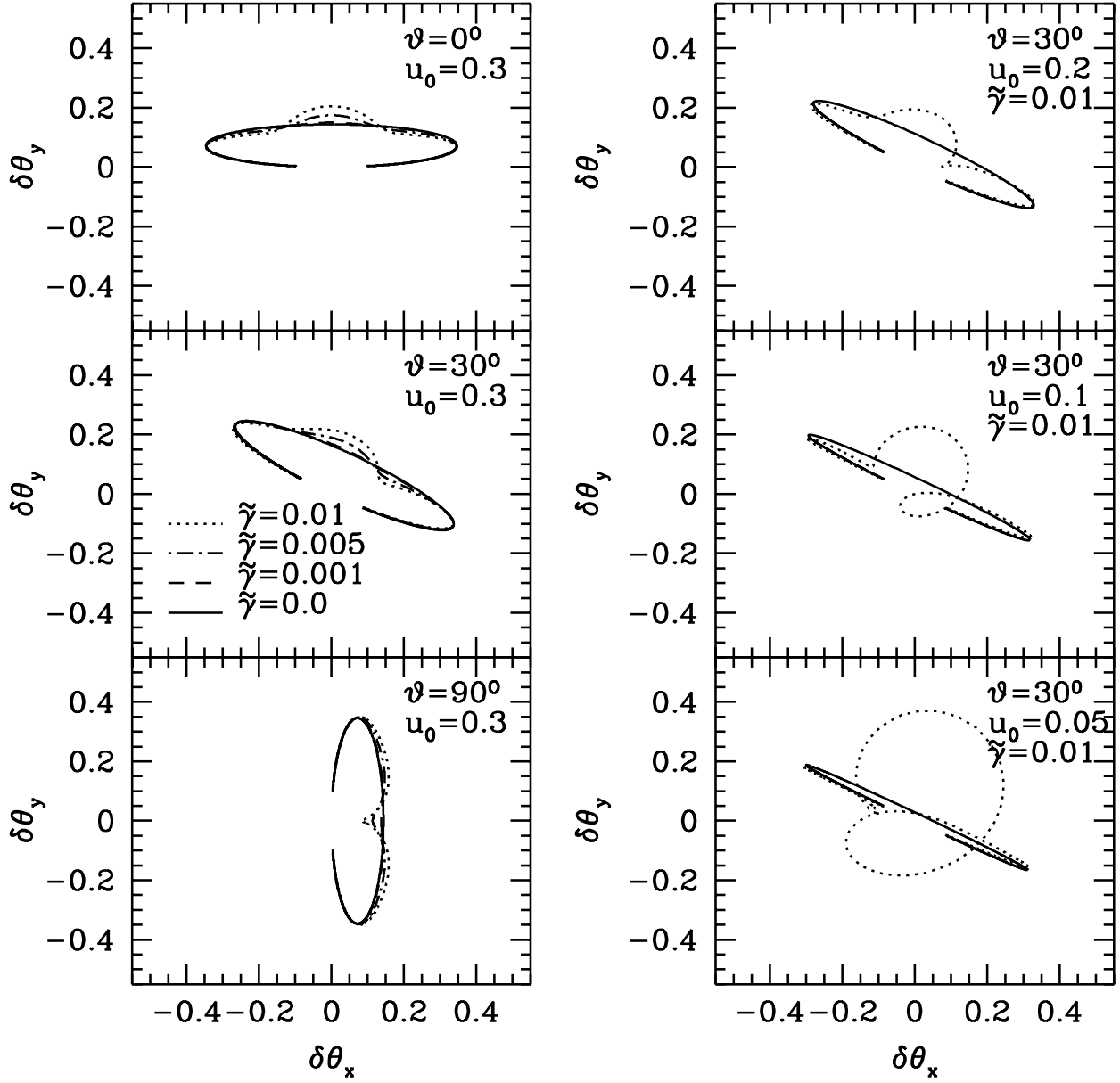}
    \caption{Centroid shift trajectories.  The left panels
    show the change of centroid shift trajectories in units of $\theta_E$ for different angle $\vartheta$ and
    shear $\gamma$, while the right panels show the variation of centroid shift trajectories for different
    impact parameter $u_0$.  The solid curves represent the simple
    astrometric ellipses.}
\end{figure}

\begin{figure}
    \epsscale{1.0} \plotone{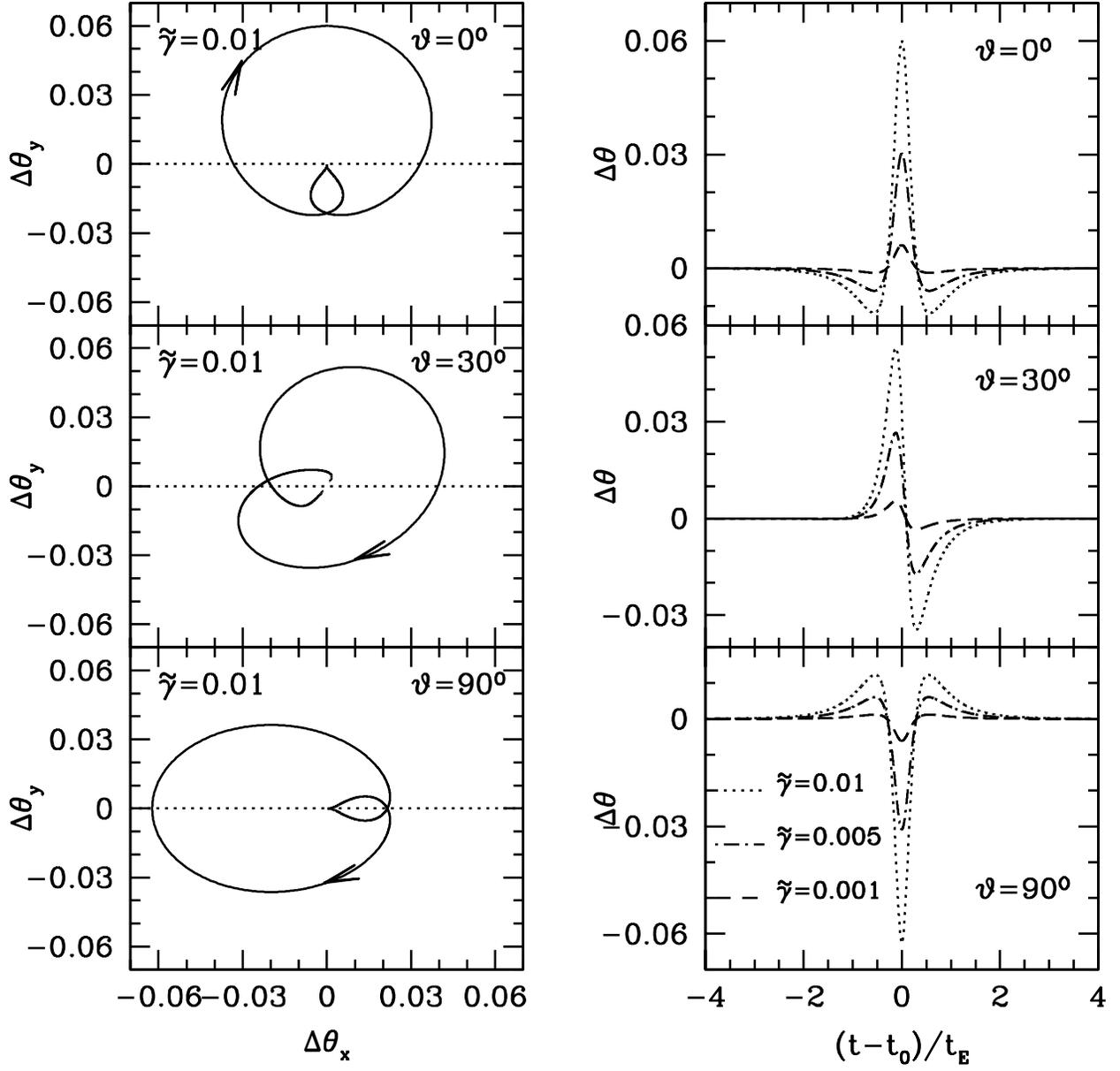}
    \caption{The trajectories and magnitudes of excess centroid shifts.
     The left panels show the trajectories of excess centroid shifts of  the events
    for $\tilde{\gamma} = 0.01$ cases in the left panels of Figure 5.   The arrow mark shows the direction of event
    progress. The right panels show the magnitudes of excess centroid
    shifts in the left panels as a function of time with three different values of shear, $\tilde{\gamma}=$ 0.01, 0.005,
    0.001.}
\end{figure}

\begin{figure}
    \epsscale{1.0} \plotone{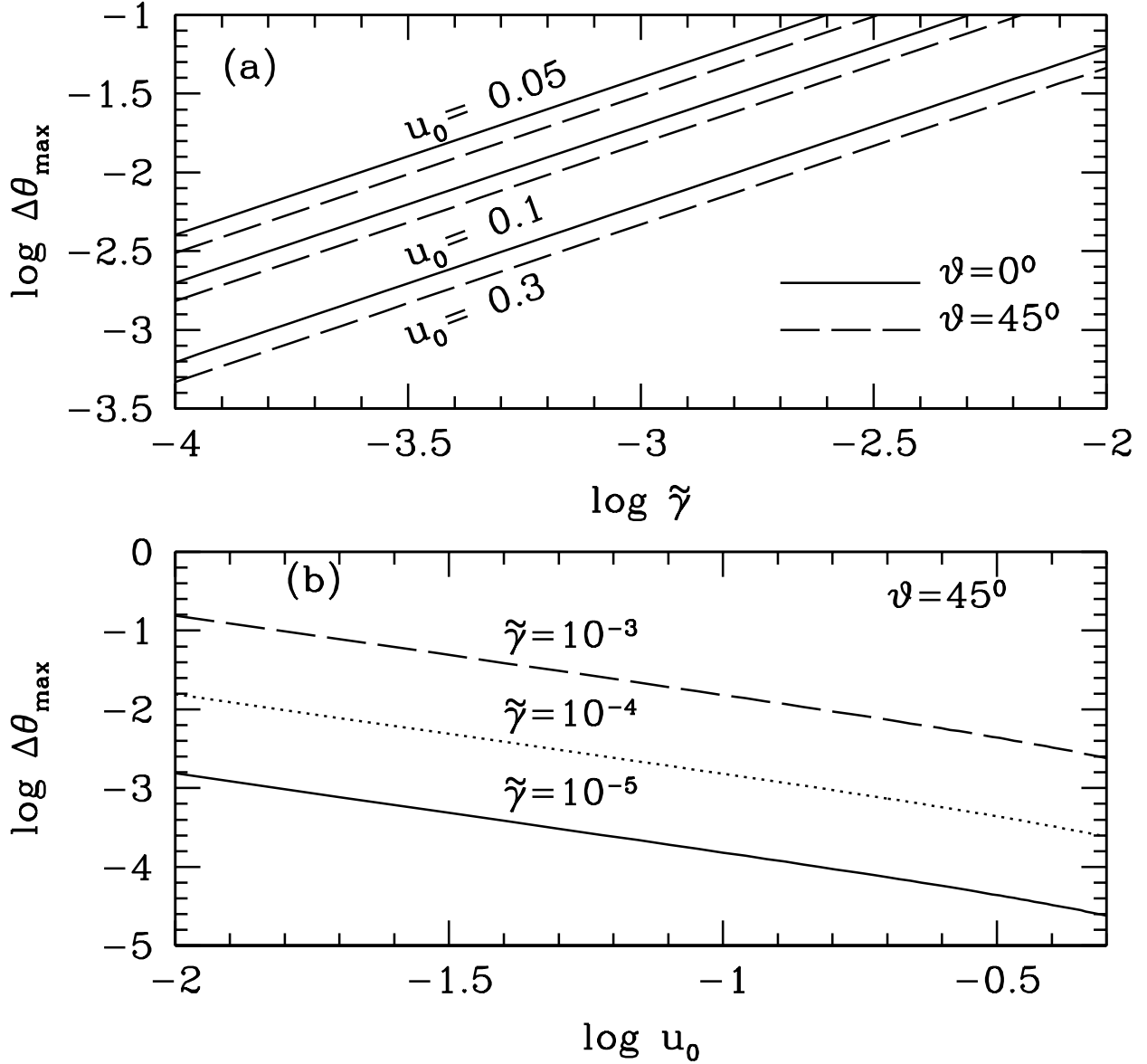}
    \caption{The maximum distortion of excess centroid shift.
    (a) The maximum distortion of excess centroid shift $\Delta\theta_{max}$
    for different values of $u_0$ and  $\vartheta$ as a function of shear $\tilde{\gamma}$.
    We show $\Delta\theta_{max}$ for $\vartheta= 0^\circ$ and $45^\circ$ only
    because other source trajectories appear between the two cases.
    (b) The maximum distortion $\Delta\theta_{max}$ as a function of $u_0$
    for different values of $\tilde{\gamma}$ and $\vartheta=45^\circ$.}
\end{figure}

\begin{figure}
    \epsscale{1.0} \plotone{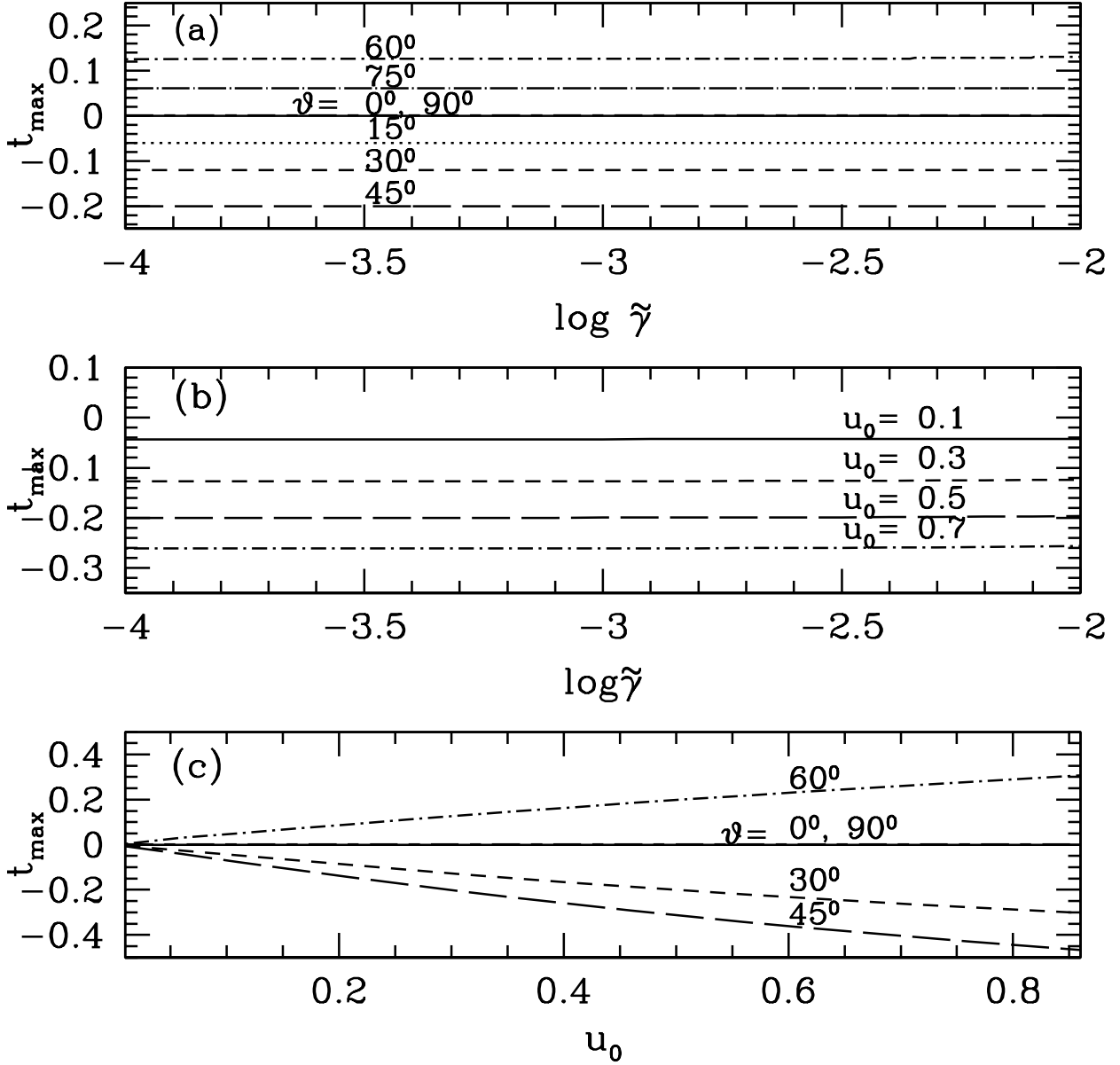}
    \caption{The time of maximum astrometric distortion.
    The time of maximum distortion of excess centroid shift,
    $t_{max}$, as a function of $\tilde{\gamma}$ and $u_0$:
    (a) for fixed $u_0=0.3$, (b) for fixed $\vartheta=30^\circ$ and (c) for fixed $\tilde{\gamma}=10^{-3}$. }
\end{figure}

\begin{figure}
    \epsscale{1.0} \plotone{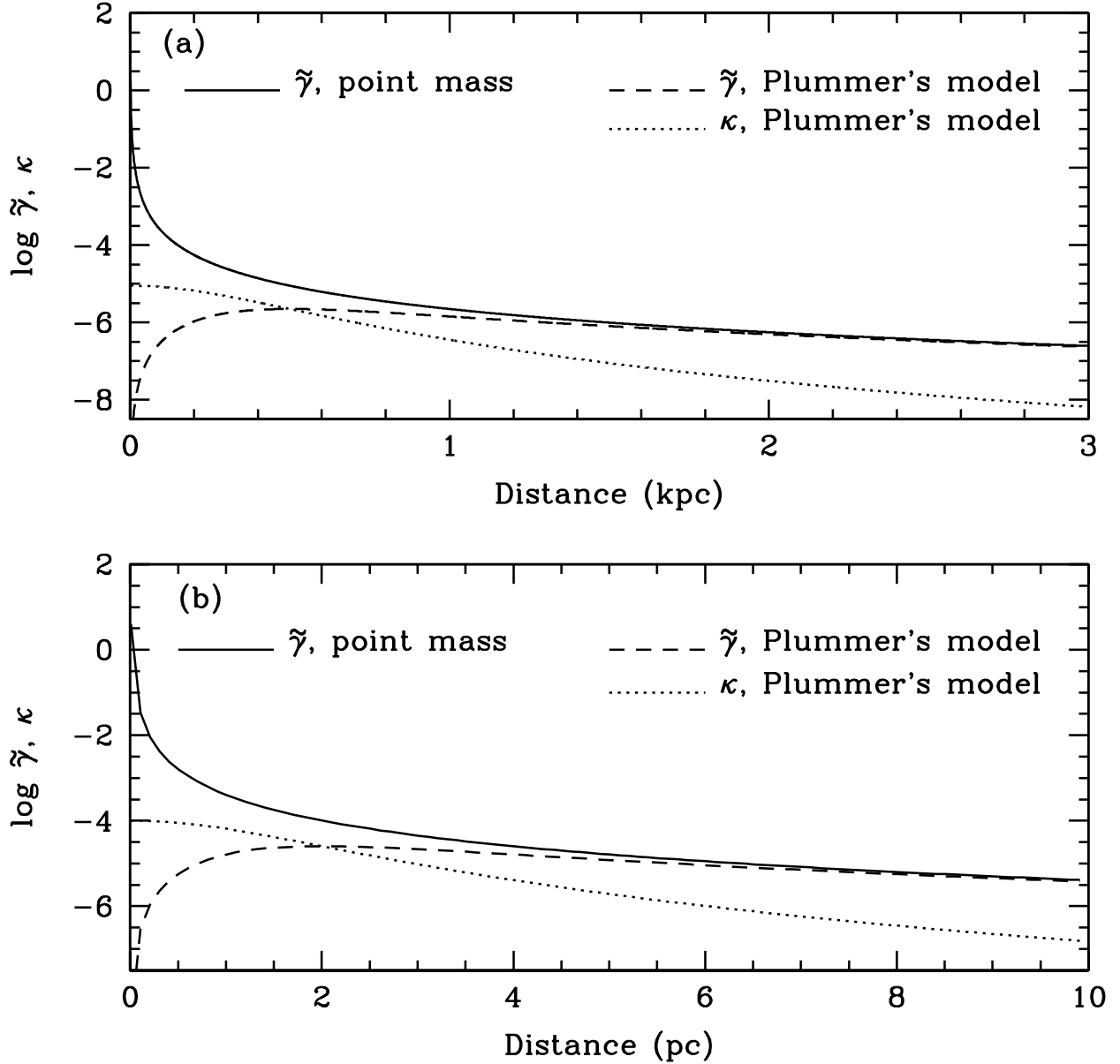}
    \caption{Shear and convergence from the Galactic sub-structures.
    (a) Shear $\tilde{\gamma}$ and convergence $\kappa$ as
    functions of the distance from the center of the mass distribution
    for $1.3\times10^{10}$ M$_\odot$ Galactic bulge at $8.5$ kpc for 1 M$_\odot$ lens at 8.5 kpc and a source star at 9.5 kpc.
    Solid curve shows the value of $\tilde{\gamma}$ when the bulge is modeled as a
    point mass while
    dotted curve as a Plummer's model with  $r_0=500$ pc. Dashed curve
    shows $\kappa$ for Plummer's model. (b) The same for
    a $10^6$ M$_\odot$ globular cluster at $4$ kpc with $r_0=2$ pc and a
    source star at $8.5$ kpc.}
\end{figure}

\begin{figure}
    \epsscale{1.0} \plotone{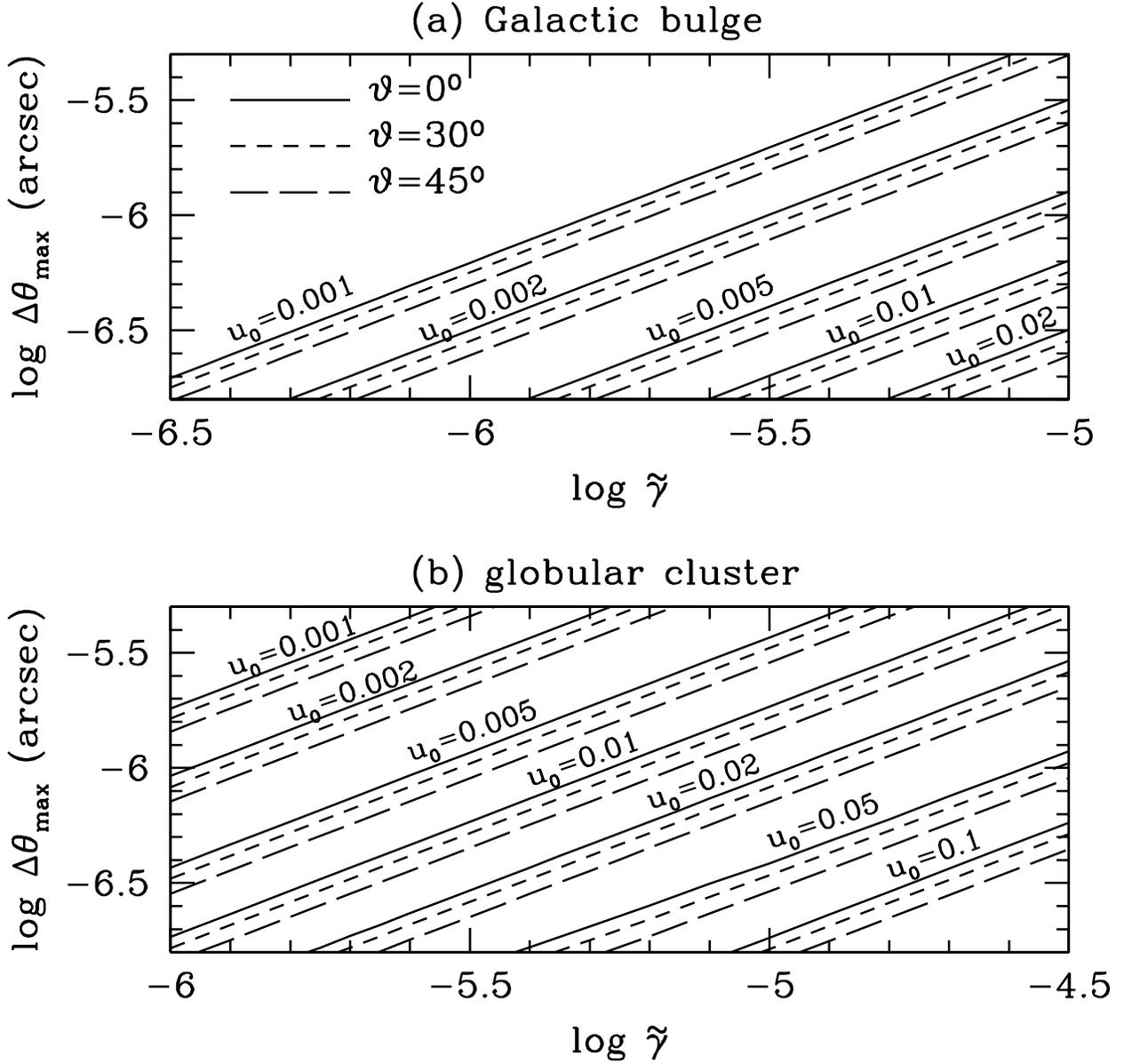}
    \caption{The maximum astrometric distortion expected.
    The maximum distortion $\Delta\theta_{max}$ in arcseconds in excess centroid
    shift as a function of the shear $\tilde{\gamma}$ due to
    (a) the Galactic bulge ($D_s=9.5$ kpc, $D_{l}=8.5$ kpc) and (b)
    a globular cluster ($D_s=8.5$ kpc, $D_{l}=4.0$ kpc), both
    with a 1 M$_\odot$ lens. Solid lines are for $\vartheta=0^\circ$, dashed ones for $\vartheta=30^\circ$,
    and long dashed ones for $\vartheta=45^\circ$. }
\end{figure}

\begin{figure}
    \epsscale{1.0} \plotone{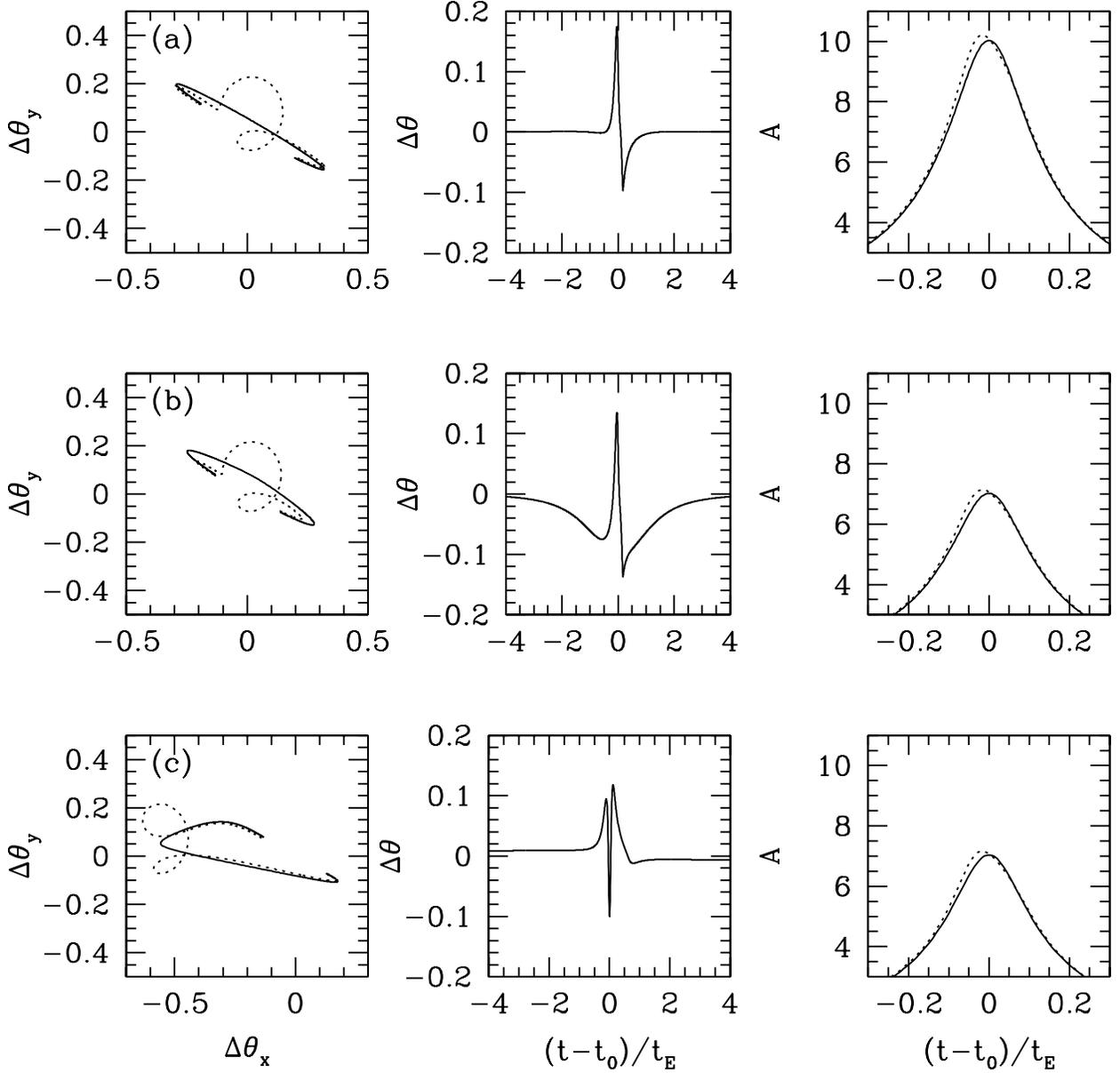}
    \caption{Complications in the centroid shift trajectory and
    the light curve from blending.
    The centroid shift trajectories (left), centroid shift deviations (middle), and light curves (right)
    are presented in case of (a) no blending (top), (b) bright lens (middle), and (c) binary
    source (bottom). The light fraction of the blending source is 0.5 for the lensing
    parameters $u_0=0.1$, $\vartheta=30^\circ$, and $\tilde\gamma=0.01$.}
\end{figure}

\end{document}